\documentclass[a4paper,UKenglish,autoref,11pt]{article}

\usepackage{microtype}

\makeatletter
\renewcommand{\maketitle}{\bgroup\setlength{\parindent}{0pt}
\begin{flushleft}
  {\huge\@title}
  
  \vspace{.5cm}
  \@author
\end{flushleft}\egroup
}
\makeatother

\usepackage{soul}

\title{Functional lower bounds for restricted arithmetic circuits of depth four}

\author{{Suryajith Chillara} \href{mailto:suryajith@cmi.ac.in}{\faEnvelopeO}\\{\scriptsize University of Haifa, Israel.}}

\date{}

\usepackage[dvipsnames,svgnames,x11names,hyperref]{xcolor}
\usepackage[framemethod=TikZ]{mdframed}
\usepackage{tikz,epigraph,verbatim,thmtools,thm-restate,ccicons,setspace,microtype}
\usepackage{array}
\usepackage{amsmath}
\usepackage{amssymb}
\usepackage{amsthm}
\usepackage{fullpage}

\usepackage{todonotes}
\usepackage{libertine}
\usepackage{eulervm}
\usepackage[T1]{fontenc}
\usepackage{inconsolata}
\usepackage{fontawesome}

\definecolor{bloodRed}{RGB}{92,3,37}
\definecolor{slightPurple}{RGB}{166,6,66}
\definecolor{slighterPurple}{RGB}{184,43,87}
\usepackage[colorlinks,backref,pagebackref]{hyperref}
\hypersetup{
  linkcolor=slightPurple, 
  citecolor=slighterPurple, 
  urlcolor= bloodRed 
}

\usepackage[capitalise]{cleveref}
\allowdisplaybreaks
\usepackage{multicol}

\newtheorem{theorem}{Theorem}
\newtheorem{corollary}[theorem]{Corollary}
\newtheorem{lemma}[theorem]{Lemma}

\newtheorem{proposition}[theorem]{Proposition}
\newtheorem{definition}[theorem]{Definition}
\newtheorem{claim}[theorem]{Claim}

\newcommand{\inparen }[1]{\left(#1\right)}             
\newcommand{\inbrace }[1]{\left\{#1\right\}}           

\newcommand{\abs}[1]{\left|#1\right|}                  

\newcommand{\fspan}{\operatorname{\F\text{-span}}}

\newcommand{\zo}{\inbrace{0,1}}                        

\newcommand{\F}{\mathbb{F}}

\newcommand{\poly}{\operatorname{poly}}

\newcommand{\NW}{\operatorname{\mathsf{NW}}}
\newcommand{\IMM}{\mathsf{IMM}}

\newcommand{\veca}{\mathbf{a}}

\newcommand{\vece}{\mathbf{e}}

\newcommand{\VP}{\mathsf{VP}}

\newcommand{\VNP}{\mathsf{VNP}}
\newcommand{\NP}{\mathsf{NP}}

\renewcommand{\P}{\mathsf{P}}

\newcommand{\FNC}{\mathsf{FNC}}

\newcommand{\ACC}{\mathsf{ACC}}

\newcommand{\MA}{\mathsf{MA}}
\newcommand{\AM}{\mathsf{AM}}
\newcommand{\PH}{\mathsf{PH}}
\newcommand{\NEXP}{\mathsf{NEXP}} 
\newcommand{\BP}[1]{\operatorname{\mathrm{BP}(#1)}}
\newcommand{\AND}{\mathsf{AND}}
\newcommand{\OR}{\mathsf{OR}}
\newcommand{\MOD}{\mathsf{MOD}}
\newcommand{\NOT}{\mathsf{NOT}}

\renewcommand{\epsilon}{\varepsilon}

\renewcommand{\epsilon}{\varepsilon}
\newcommand{\ignore}[1]{}

\newcommand{\mrc}{multi-$r$-ic\ }

\newcommand{\supp}[1]{\operatorname{\mathrm{Supp}}(#1)}
\newcommand{\monsupp}[1]{\operatorname{\mathrm{MonSupp}}(#1)}

\newcommand{\mult}{\operatorname{\mathrm{mult}}}
\newcommand{\ml}{\mathrm{ML}}

\newcommand{\psspd}{\operatorname{\mathrm{PSSPD}}_{k,\ell}^{[Y,Z]}}
\newcommand{\psspdr}{\operatorname{\mathrm{PSSPD}}_{rk,\ell}^{[Y,Z]}}

\newcommand{\sedfks}{\operatorname{\mathrm{SED}}_{k,\ell}^{[Y,Z]}}
\newcommand{\sed}{\operatorname{\mathrm{mSED}}_{k,\ell}^{[Y,Z]}}

\newcommand{\eval}{\operatorname{\mathrm{Eval}}}
\newcommand{\SPSP}{\Sigma\Pi\Sigma\Pi}

\newcommand{\SWSP}{\Sigma\mathord{\wedge}\Sigma\Pi}

\newcommand{\GapL}{\mathsf{GapL}}

\makeatletter

\begin{document}
\vskip1mm
\hrule
\vskip1mm

\maketitle

\vskip1mm
\hrule
\vskip1mm

\begin{abstract}
  Recently, Forbes, Kumar and Saptharishi [CCC, 2016] proved that there exists an explicit $d^{O(1)}$-variate and degree $d$ polynomial $P_{d}\in \VNP$ such that if any depth four circuit $C$ of bounded formal degree $d$ which computes a polynomial of  bounded individual degree $O(1)$, that is functionally equivalent to $P_d$, then $C$ must have size $2^{\Omega(\sqrt{d}\log{d})}$.

  The motivation for their work comes from Boolean Circuit Complexity. Based on a characterization for $\ACC^0$ circuits by Yao [FOCS, 1985] and Beigel and Tarui [CC, 1994], Forbes, Kumar and Saptharishi [CCC, 2016] observed that functions in $\ACC^0$ can also be computed by algebraic $\SWSP$ circuits (i.e., circuits of the form -- sums of powers of polynomials) of $2^{\log^{O(1)}n}$ size. Thus they argued that a $2^{\omega\inparen{\poly\log{n}}}$ ``functional'' lower bound for an explicit polynomial $Q$ against $\SWSP$ circuits would imply a lower bound for the ``corresponding Boolean function'' of $Q$ against non-uniform $\ACC^0$. In their work, they ask if their lower bound be extended to $\SWSP$ circuits.
  
  In this paper, for large integers $n$ and $d$ such that $\omega(\log^2n)\leq d\leq n^{0.01}$, we show that any $\SWSP$ circuit of bounded individual degree at most $O\inparen{\frac{d}{k^2}}$ that functionally computes Iterated Matrix Multiplication polynomial $\IMM_{n,d}$ ($\in \VP$) over $\zo^{n^2d}$ must have size $n^{\Omega\inparen{k}}$. Since Iterated Matrix Multiplication $\IMM_{n,d}$ over $\zo^{n^2d}$ is functionally in $\GapL$, improvement of the afore mentioned lower bound to hold for quasipolynomially large values of individual degree would imply a fine-grained separation of $\ACC^0$ from $\GapL$.

  For the sake of completeness, we also show a syntactic size lower bound against any $\SWSP$ circuit computing $\IMM_{n,d}$ (for the same regime of $d$) which is tight over large fields. Like Forbes, Kumar and Saptharishi [CCC, 2016], we too prove lower bounds against circuits of bounded formal degree which functionally compute $\IMM_{n,d}$, for a slightly larger range of individual degree.
\end{abstract}

\section{Introduction}

Owing to the difficulty in proving Boolean circuit size lower bounds, Valiant proposed that we prove lower bounds in an ``algebraic setting'' as the underlying algebraic structure could help us understand the computations better. Valiant further conjectured that any circuit theoretic proof for $\P \neq \NP$ would have to be preceded by an analogous result in this more constrained arithmetic model \cite{v92}.

Arithmetic circuits (also called as algebraic circuits) are directed acyclic graphs such that the leaf nodes are labeled by variables or constants from the underlying field, and every non-leaf node is labeled either by a $+$ or $\times$. Every node computes a polynomial by  operating on its inputs with the operation given by its label. The computation flows from the leaves to the output node. Complexity of computation here is quantified by the size of the circuit, which is the number of nodes in it.



It is conjectured that Permanent polynomial does not have polynomial size arithmetic circuits \cite{val79}. B\"{u}rgisser~\cite{bur00} showed that if Permanent polynomial were to have a polynomial sized arithmetic circuit then this would imply $\#\P\subseteq \FNC^3/\poly$  which would further imply that $\NP\subseteq \P/\poly$ which leads to (1) $\PH\subseteq \Sigma^2_p$ \cite{KL80} and (2) $\AM=\MA$ \cite{AKSS}, both of which go against widely believed conjectures. Thus, a central question in the field of algebraic complexity theory is to show that Permanent polynomial (or any closely related polynomial of interest) needs superpolynomial sized arithmetic circuits to compute it.

Four decades after the problem was formulated, the best known size lower bound is still super linear~\cite{BS}. Over the span of last three decades, researchers have considered restricted arithmetic circuits and here we have seen a great progress towards proving lower bounds under these restrictions (see \cite{sy, github} for a detailed survey). In a surprising result, Agrawal and Vinay \cite{av} showed that it is sufficient to prove subexponential size lower bounds against depth four circuits, to prove super polynomial size lower bounds against general arithmetic circuits.

A depth four circuit\footnote{Generally speaking, a depth four circuit can also be of the form $\Pi\Sigma\Pi\Sigma$ but we follow the convention that the root node is a $+$ node. Under such a convention $\Pi\Sigma\Pi\Sigma$ circuit is a depth five circuit.} (denoted by $\SPSP$) computes polynomials that can also be expressed as a sum of products of polynomials.
  \begin{align*}
    P(X) = \sum_{i=1}^{s_1}\prod_{j} Q_{i,j}\,.
  \end{align*}
  \paragraph*{Syntactic lower bounds:} We say that a polynomial $P$ has a syntactic circuit size lower bound of $s$ against class $\mathcal{C}$ of circuits if no circuit in $\mathcal{C}$ of size strictly smaller than $s$ syntactically computes $P$.

  Strong syntactic size lower bounds for depth four circuits were proven in restricted settings: Bounded fan-in \cite{GKKS, KSS, FLMS, CM19, KStopfanin}, Homogeneous \cite{KLSS14a, KS14prelim, KLSS, KS14}, Multilinear \cite{ry09, CLS}, and Multi-$r$-ic \cite{KST, ChiSTACS20, ChiExp20}. In a breakthrough, Limaye, Srinivasan and Tavenas recently proved superpolynomial size lower bounds against all constant depth circuits \cite{LST}. Prior to that the best known lower bound for depth four circuits was super-quadratic \cite{GST} (which improves upon super-linear lower bounds due to Shoup and Smolensky~\cite{SS97} and Raz \cite{Raz10}).

  \paragraph*{Functional lower bounds:} For a set $B\subseteq \F$, we say that two polynomials $P(x_1,\ldots, x_N)$ and $Q(x_1, \ldots, x_N)$ are functionally equivalent over $B^N$ if $P(\veca) = Q(\veca)$ for all $\veca\in B^N$. We say that a circuit $C$ functionally computes a polynomial $P\in \F[x_1,\ldots, x_N]$ over $B^N$ if the output polynomial $f\in F[x_1,\ldots, x_N]$ of $C$ is functionally equivalent to $P$ over $B^N$.

  We say that a polynomial $P$ has a functional size lower bound of $s$ against a class $\mathcal{C}$ of circuits if no polynomial that is computed by circuits in $\mathcal{C}$ of size strictly less than $s$, is functionally equivalent to $P$ over $B^n$ for any $B\subseteq \F$.

  Forbes, Kumar and Saptharishi \cite{FKS16} proved exponential functional lower bounds for a polynomial in $\VNP$ against depth four circuits of bounded formal degree and bounded individual degree $O(1)$. Formally, they showed that there is an explicit polynomial $P_{d}$ of degree $d$ over $\approx d^3$ variables such that no depth four circuit of bounded formal degree $d$ and size smaller than $2^{c(\sqrt{d}\log{d})}$ (for a small constant $c$) that computes a polynomial of bounded individual degree at most $O(1)$  can be functionally equivalent to $P_{d}$. Apart from this work, strong functional lower bounds are known against depth three circuits over finite fields \cite{GR00}, multilinear formulas \cite{Raz, Raznc2nc1, ry08, ry09, CELS, CLS}, and set-multilinear formulas \cite{NW97, LST}.

  The motivation for the work of \cite{FKS16} comes from Boolean circuit complexity. $\ACC^0$ circuits are constant depth Boolean circuits that have $\AND$, $\OR$, $\NOT$ and $\MOD$ gates. Allender and Gore \cite{AG94} showed that \emph{uniform} $\ACC^0$ circuits of subexponential size cannot compute Permanent. In a major breakthrough, Williams \cite{Wil14} showed that there exists a function in $\NEXP$ such that it cannot be computed by polynomial sized \emph{nonuniform} $\ACC^0$ circuits. Recently Murray and Williams~\cite{MW20} further improved the situation to show that there exists a function in $\mathsf{NQP}$ such that it needs superpolynomial size $\ACC^0$ circuits to compute it. 

  Beigel and Tarui \cite{BT94} showed that every language $L$ in the class $\ACC^0$ can be recognized by a family of depth two\footnote{Here the variables can appear negated at the leaves that feed into the $\AND$ gates. Even though it is stated as depth two in the paper, the longest leaf to root path in this circuit is of length 3.  $\text{Leaf node}~\rightarrow~\AND~\rightarrow~\text{root}$.} deterministic circuits with a symmetric function gate at the root and $2^{\log^{O(1)}{n}}$ many $\AND$ gates of fan-in $\log^{O(1)}{n}$ in the second layer. Over large fields, Forbes, Kumar and Saptharishi~\cite{FKS16} observed that given this Boolean circuit, there is an algebraic circuit of depth four which computes polynomials of the form -- sum of $2^{\log^{O(1)}{n}}$ many powers of polynomials each of whose monomials are supported on at most $\log^{O(1)}n$ many variables such that outputs of both of these circuits are functionally equivalent.

  $\SWSP$ circuits are depth four circuits that compute polynomials which can be expressed as sums of powers of polynomials. $\SWSP^{[t]}$ circuits are depth four circuits that compute polynomials which can be expressed as sums of powers of polynomials each of whose monomials are supported on at most $t$ many variables.

  We can summarize the afore mentioned discussion formally as follows.
  \begin{lemma}[Lemma 3.2, \cite{FKS16}]
    Let $\F$ be any field of characteristic zero or at least $\exp(\omega(\poly(\log{n})))$. If a function $f:\zo^n\mapsto\zo$ is in $\ACC^0$ then there exists a polynomial $P_f\in \F[x_1,\ldots, x_n]$ such that
    \begin{itemize}
    \item $P_f$ and $f$ are functionally equivalent over $\zo^n$, and
    \item $P_f$ can be computed by a $\SWSP$ circuit of top fan-in at most $2^{\log^{O(1)}n}$ and bottom support at most $\log^{O(1)}n$.
    \end{itemize}
  \end{lemma}

  Thus, to show a lower bound against $\ACC^0$ circuits in the Boolean setting, it is sufficient to show a functional lower bound of $\exp(\omega(\poly(\log{n})))$ for a polynomial $P$ would imply that the Boolean part\footnote{B\"{u}rgisser \cite{bur00} defined the boolean part of a polynomial $P(x_1,\ldots, x_n)$ (denoted by $\BP{P}$) to be a function that agrees with $P$ over all evaluations over $\zo^n$.} of $P$ is not in $\ACC^0$.
  
  \begin{lemma}[Lemma 3.3, \cite{FKS16}]
    Let $\F$ be any field of characteristic zero or at least $\exp(\omega(\poly(\log{n})))$. Then a $\exp(\omega(\poly(\log{n})))$ functional size lower bound for a $n^{O(1)}$-variate and $n^{O(1)}$ degree polynomial $P\in \F[X]$ against $\SWSP^{[\poly(\log(n))]}$ circuits over $\F$ would imply that Boolean part of $P$ is not in $\ACC^0$.
  \end{lemma}

  Forbes, Kumar and Saptharishi~\cite{FKS16} through an open question in their paper ask if such functional lower bounds can also be proved for $\SWSP$ circuits. We in this paper show strong functional lower bounds against all $\SWSP$ circuits which output polynomials of bounded individual degree.

  A circuit $C$ is said to have a bounded individual degree\footnote{Not to be confused with the \mrc circuits dealt with in \cite{KS17, KST, ChiSTACS20, ChiExp20}.} $r$ if the polynomial output by the circuit $C$ has degree at most $r$ with respect to each of its variables.

\begin{theorem}[Functional Lower Bounds for $\SWSP$ circuits of Bounded Individual Degree]
  \label{thm:fn-pow-d4}
  Let $n$ be a large integer. Let $d, k$ and $r$ be such that $\omega(\log^2n)\leq d \leq n^{0.01}$ and $r\leq \frac{d}{1201k^2}\,.$ Any depth four $\SWSP$ circuit of bounded individual degree $r$ computing a function equivalent to $\IMM_{n,d}$ on $\zo^{n^2d}$, must have size at least $n^{\Omega\inparen{k}}$.
\end{theorem}

Note that there is a trade-off between the lower bound on the circuit size and the upper bound on the range of $r$ this lower bound can be achieved for.

Since Iterated Matrix Multiplication $\IMM_{n,d}$ over $\zo^{n^2d}$ is functionally\footnote{B\"{u}rgisser \cite{bur00} showed that boolean part of any polynomial in $\VP$ lies in $\FNC^3/\poly$, and in particular $\IMM_{n,d}\in \VP$. On the other hand, Vinay~\cite{Vin91} identified that this problem of computing Iterated Matrix Product of integer matrices (denoted by $\mathrm{ITMATPROD}$) is in fact in the class $\GapL$ which consists of all problems that are logspace reducible to determinant computation of an integer matrix. This is a better characterization as $\GapL\subseteq NC^2\subseteq \FNC^3/\poly$.} in $\GapL$ \cite[Section 6]{Vin91}, improvement of the afore mentioned lower bound to hold for quasipolynomially large values of individual degree would imply a fine-grained separation of $\ACC^0$ from $\GapL$.

By a divide and conquer construction, we get a depth four $\SPSP$ circuit of size $n^{O(\sqrt{d})}$ that computes $\IMM_{n,d}$ such that the fan-in of both the product gates is equal to $\sqrt{d}$. Using the identity 
\begin{align*}
  m! \cdot x_1x_2\ldots x_m = \sum_{S\subseteq [m]}\inparen{\sum_{i\in S} x_i}^m\cdot(-1)^{m-\abs{S}}
\end{align*}
(attributed to Fischer \cite{Fis} and Ryser \cite{Rys} in \cite{GKKSdepth3}), over large fields this circuit can be converted into a $\SWSP$ circuit of size $n^{O(\sqrt{d})}$. We will now show a lower bound of $n^{\Omega(\sqrt{d})}$ for $\IMM_{n,d}$ against any $\SWSP$ circuits. From the afore mentioned discussion, this lower bound is optimal up to a   constant in the exponent over large fields.

\begin{theorem}[Syntactic Lower Bounds for $\SWSP$ circuits]
  \label{thm:syn-pow-d4}
  Let $n$ and $d$ be a large integers such that $\omega(\log^2n)\leq d \leq n^{0.01}$. Any depth four $\SWSP$ circuit computing  $\IMM_{n,d}$ must have size at least $n^{\Omega\inparen{\sqrt{d}}}$.
\end{theorem}

Recall that Forbes, Kumar and Saptharishi \cite{FKS16} proved functional lower bounds for a polynomial in $\VNP$ against depth four circuits of bounded formal degree whose output polynomials are of bounded individual degree $O(1)$. Here shall prove functional lower bounds for a polynomial in $\VP$ against depth four circuits of bounded formal degree whose output polynomials are of bounded individual degree $O(\log{n})$.

Formal degree of a circuit is the maximum degree of any polynomial that could be computed by this circuit structure sans the constants nor cancellations. 
Formal degree of a circuit is inductively defined as follows: for a leaf node $w$, the formal degree $1$ if it is labeled by a variable and $0$ otherwise. Formal degree of a sum node is the maximum over all the formal degrees of its children, and formal degree of a product node is equal to the sum over all the formal degrees of its children. 

\begin{theorem}[Functional Lower Bounds for $\SPSP$ Circuits of Bounded Formal Degree]
  \label{thm:fn-prod-d4}
  Let $n$, $d$ and $r$ be integers such that $\Omega(\log^2n)\leq d \leq n^{0.01}$ and $r\leq \frac{\log{n}}{12}\,.$ Any depth four $\SPSP$ circuit of formal degree $d $ and bounded individual degree $r$ that computes a function equivalent to $\IMM_{n,d}$ on $\zo^{n^2d}$, must have size at least $n^{\Omega\inparen{\sqrt{\frac{d }{r}}}}$.
\end{theorem}

We would to remark that the afore mentioned bound and the bound for similar circuits in \cite{FKS16} can be made to work for formal degree that is slightly larger than $d$. 

\subsection*{Related Work}

For the sake of brevity, we shall denote the $\SWSP$ circuits of bounded individual degree $r$ by $(\SPSP)^{\leq r}$. We in this table summarize our results in comparison to the work of \cite{FKS16}.
{\small
\begin{center}
  \begin{tabular}{|m{0.15\textwidth}|m{0.1\textwidth}|m{0.2\textwidth}|c|m{0.25\textwidth}|}
  \hline
    Circuit model & Work & Hard multilinear polynomial family &  Lower Bound & Range of parameters\\
    \hline\vskip2mm
    $(\SPSP)^{\leq r}$ \& formal degree $d$ & \cite{FKS16} & Nisan-Wigderson polynomial $\NW_{m,d}\in\VNP$ with $md$ many variables and degree $d$  &  $ {2}^{\Omega\inparen{\sqrt{d}\log{(md)}}}$& $ m=\Theta(d^2)$, and $r\leq O(1)$.\\
    \hline\vskip2mm
    $(\SPSP)^{\leq r}$ \& formal degree $d$ & This work & Iterated Matrix Multiplication polynomial $\IMM_{n,d}\in\VP$ with $n^2d$ many variables and degree $d$ &  $ {n}^{\Omega\inparen{\sqrt{\frac{d}{r}}}}$& $ \omega(\log^2{n})\leq d\leq n^{0.01}$, and $r\leq \frac{\log{n}}{12}$.\\
  \hline
    $(\SWSP)^{\leq r}$ & This work & $\IMM_{n,d}$ &  $ {n}^{\Omega\inparen{k}}$& $ \omega(\log^2{n})\leq d\leq n^{0.01}$, and $r\leq \frac{d}{1201k^2}$.\\
  \hline
\end{tabular}
\end{center}
}
Our work is inspired by \cite{FKS16}'s line of research and depends on the techniques introduced by them. We take their research a bit further.

\subsection*{Complexity measure and proof overview}
Let the variable set $X$ be partitioned into two fixed, disjoint sets $Y$ and $Z$. Let $\sigma_Y: \F[Y\sqcup Z] \mapsto \F[Z]$ be a linear map such that for any polynomial $P(Y,Z)$, $\sigma_Y(P)\in \F[Z]$ is obtained by setting every variable from $Y$ to zero and leaving the variables from $Z$ untouched.


For a polynomial $P(x_1,\ldots, x_N)$, let $\mult(P)$ be defined to be equal to $P\mod \inbrace{(x_i^2 - x_i) \mid i\in[N]}$. Similarly, let $\mult(V)$ for a subspace $V$ of polynomials in  $\subseteq \F[x_1,\ldots,x_N]$, be defined as follows.
\begin{align*}
  \mult(V) = \inbrace{\mult(P)\mid P\in V}.
\end{align*}

For a polynomial $P(Y,Z)$ and a set $S\subseteq \F$, let $\eval_S^{[Y\cup Z]}(P)$ denote the vector of evaluations of polynomial $P$ over $S^{\abs{Y\cup Z}}$ as follows.
\begin{align*}
  \eval_S^{[Y\cup Z]}(P(Y,Z)) = (P(\veca))_{\veca \in S^{\abs{Y\cup Z}}}\,.
\end{align*}
This definition can be extended to a set $V$ of polynomials over $\F[Y\cup Z]$ as follows.
\begin{align*}
  \eval_S^{[Y\cup Z]}(V) &= \inbrace{\eval_S^{[Y\cup Z]}(P(Y,Z))\mid P(Y,Z) \in V}\,.
\end{align*}

We use $\partial_Y^{\leq k}{P}$ to denote the set of all partial derivatives of $P$ of order at most $k$ with respect to monomials over variables just from $Y$, and  $Z^{= \ell}\cdot \sigma_Y(\partial_Y^{=k}P)$ to refer to the set of polynomials obtained by multiplying each polynomial in $\sigma_Y(\partial_Y^{\leq k}P)$ with monomials of degree equal to $\ell$ in $Z$ variables.



\paragraph*{Main measure -- Multilinear Shifted Evaluation Dimension ($\sed$):} Forbes, Kumar and Saptharishi \cite{FKS16} defined Shifted Evaluation Dimension which counts the dimension of space of vectors each of which is a list of evaluations of polynomials $\zo^{\abs{X}}$ where these polynomials are $Z$-shifts of partial evaluations.
\begin{align*}
  \sedfks(P(Y,Z)) = \dim\inparen{\eval_{\zo^{\abs{Z}}}\inbrace{{Z^{= \ell}\cdot \fspan\inbrace{P(\veca, Z)\mid \veca \in \zo^{\abs{Y}}_{\leq k}}}}}
\end{align*}

We just make a minor modification to this measure to better relate our measure with the measure of Projected Shifted Skew Partial derivatives (\cite{ChiSTACS20, ChiExp20}) and this helps us obtain bounds that we could not get before.
\begin{align*}
  \sed(P(Y,Z)) = \dim\inparen{\eval_{\zo^{\abs{Z}}}\inbrace{\mult\inparen{Z^{= \ell}\cdot \fspan\inbrace{P(\veca, Z)\mid \veca \in \zo^{\abs{Y}}_{\leq k}}}}}
\end{align*}
In spirit, it is still the measure of \cite{FKS16} and thus we do not consider this to be a new measure. We just make a minor modification to relate this measure with their measure of Projected Shifted Skew Partial derivatives (\cite{ChiSTACS20, ChiExp20}) and this helps us obtain bounds that we could not get before. 

By unfurling the above definition, we can see that if two $N$-variate polynomials $P_1(Y,Z)$ and $P_2(Y,Z)$ (defined on the same variable sets) are functionally equivalent over $\zo^N$ then $\sed(P_1(Y,Z))= \sed(P_2(Y,Z))$. Note that two polynomials which are not functionally equivalent over $\F^N$ can end up being functionally equivalent over $\zo^N$ but to show that two polynomials are not functionally equivalent, it is sufficient to show that they are not functionally equivalent over $\zo^N$.

The crux of our work henceforth is to show that the polynomial of interest, $\IMM_{n,d}$ is not functionally equivalent over $\zo^{n^2d}$ to the polynomials that are output by the $\SWSP$ circuits of bounded individual degree. That is, we need to show that $\sed(\IMM_{n,d}(Y,Z))$ is much larger than $\sed(C(Y,Z))$ where $C$ is a $\SWSP$ circuit of small size and bounded individual degree.

Though two $N$-variate polynomials $P_1$ and $P_2$ that are functionally equivalent over $\zo^N$ have the same (multilinear) shifted evaluation dimension, the dimension of their partial derivative spaces can be very different (see \cite[Section 1.2.1]{FKS16} for an example). However in certain special cases Forbes, Kumar and Saptharishi~\cite{FKS16} do manage to relate the shifted evaluation dimension, and a partial derivate based measure well enough for their proof to work. We shall do something very similar.

Let $C$ be a $\SWSP$ circuit of bounded individual degree at most $r$ that computes a polynomial that is functionally equivalent to a homogeneous and degree $d$ set-multilinear polynomial $P(X)$ defined over the sets $X=X_1\sqcup\ldots\sqcup X_d$ such that $Y = X_{i_1}\sqcup\ldots X_{i_k}$ (for a fixed subset $\inbrace{i_1, \ldots, i_k}\subseteq [d]$) and $Z=X\setminus Y$.
Similar to \cite{FKS16}, we show that we can bound the multilinear shifted evaluation dimension on the above and below by an auxiliary measure that counts the dimension of a space of a specially chosen syntactic polynomials. For every value of $k$, $\ell$ and $r$, we can show that
\begin{align*}
  \psspd(P(Y,Z)) \leq \sed(P(Y,Z)) =  \sed(C(Y,Z))\leq \psspdr(C(Y,Z))\,.
\end{align*}

Upon instantiating the above expression with explicit homogeneous and set-multilinear polynomial $\IMM_{n,d}(Y,Z)$, and if for a suitable setting of values of $k,\ell$ and $r$, we get that $\psspd(\IMM_{n,d}(Y,Z))$ is much larger than $\psspdr(C(Y,Z))$ where $C$ is a $\SWSP$ circuit that computes polynomials of bounded individual degree $r$ of size $s$, then we can infer that $\IMM_{n,d}(Y,Z)$ cannot be functionally computed by this class of circuits, thus giving us a functional size lower bound of $s$ for this explicit polynomial. 

\paragraph*{Auxiliary measure -- Projected Skew Shifted Partial Derivatives ($\psspd$):}
The following is a measure\footnote{This measure is an amalgamation of measures -- dimension of Projected Shifted Partial derivatives of \cite{KLSS14a} and dimension of Skew Shifted Partial derivatives of \cite{KST}.} borrowed from \cite{ChiExp20} which was used to prove syntactic lower bounds for \mrc depth four circuits. 
\begin{align*}
  \psspd(P(Y,Z)) = \dim\inparen{\F\text{-span}\inbrace{\mult\inparen{Z^{= \ell}\cdot \sigma_Y\inparen{\partial^{\leq k}_Y P}}}}\,.
\end{align*}

We currently do not know how to directly obtain a bound on $\psspdr(C(Y,Z))$ to a value that is much smaller than $\psspd(\IMM_{n,d}(Y,Z))$. To resolve this issue, we use random restrictions $V\leftarrow D$ to convert our $\SWSP$ circuit $C$ of size $s\leq n^{\frac{t}{2}}$ that computes a polynomial $P$ of bounded individual degree to a $\SWSP$ circuit $C'$ of size $s$ and of bottom fan-in at most $t$ that still computes the restricted polynomial $P'$, with a high probability. We can now bound $\psspdr(C(Y,Z))$ to a value that is much smaller than $\psspd((\IMM_{n,d}(Y,Z))|_V)$. This trick is omnipresent in this line of work \cite{KLSS14a, KS14prelim, KLSS, KS14, KST, FKS16, ChiSTACS20, ChiExp20}.

We then borrow the lower bound on $\psspd(P'(Y,Z))$ (where $P'$ is the polynomial obtained from $\IMM_{n,d}$ after restrictions) from \cite{ChiExp20}.

We would like to remark that $\mult(P)$ for a polynomial $P(x_1,\ldots,x_N)$ was defined to be $P\mod \inbrace{x_i^2:i\in[N]}$ in \cite{ChiSTACS20, ChiExp20} instead of $P\mod \inbrace{x_i^2-x_i:i\in[N]}$ as defined here.
We use this new definition of $\mult$ because $\sed(P_1(Y,Z))$ may not be equal to $\sed(P_2(Y,Z))$ under the older definition of $\mult(P) = P\mod \inbrace{x_i^2:i\in[N]}$ even though $P_1(Y,Z)$ and $P_2(Y,Z)$ are functionally equivalent.

The lower bound on $\psspd(P'(Y,Z))$ in \cite{ChiExp20} continues to hold despite this change of definition.

\section{Preliminaries}
\paragraph*{Notation:}
\begin{itemize}
\item We use $[n]$ to refer to the set $\inbrace{1,2,\ldots, n}$.
\item For a polynomial $f$ and a monomial $m$ of degree $k$, we use $\partial^{k}_m f$ to refer to the $k$th partial derivate of the polynomial $f$ with respect to the monomial $m$.
\item For a polynomial $f$, we use $\partial^{\leq k}_Y(f)$ to refer to the space of partial derivatives of order at most $k$ of $f$ with respect to monomials of degree at most $k$ in variables from $Y$.
\item We use $Z^{=\ell}$ and $Z^{\leq \ell}$ to refer to the set of all the monomials of degree equal to $\ell$ and at most $\ell$, respectively, in variables $Z$. 
\item We use $Z_{\ml}^{\leq t}$ to refer to the set of all the multilinear monomials of degree at most $t$ in $Z$ variables.
\item For sets $A$ and $B$ of polynomials, we define the product $A\cdot B$ to be the set $\inbrace{f\cdot g\mid f\in A~\text{and}~g\in B}$.
\item For a monomial $m$ we use $\supp{m}$ to refer to the set of variables that appear in it.
\item We use $Z_{\inbrace{\leq t}}$ to refer to the set of all monomials $m$ in $Z$ variables such that $\abs{\supp{m}}\leq t$.
\end{itemize}

\begin{claim}
  \label{clm:eval-sync-mlequiv}
  Let $W\subseteq \F[X]$ be a subspace of multilinear polynomials. Then $\dim(W) = \dim (\eval_{\zo}^{[X]}(W))$. 
\end{claim}
\begin{proof}
  Proof of this claim follows from the facts that every multilinear polynomial in $W$ has a unique evaluation vector, and access to evaluations of a multilinear polynomial over all of $\zo^{\abs{X}}$ uniquely determines it.
\end{proof}

\begin{proposition}
  \label{prop:mult}
  For two sets $A$ and $B$ of polynomials,
  \begin{enumerate}
  \item $\mult(A\cdot B) = \mult(\mult(A)\cdot\mult(B))$, and
  \item $\dim(\mult(\mult(A)\cdot\mult(B))) \leq \dim(\mult(A)\cdot\mult(B))$.
  \end{enumerate}
\end{proposition}
The proof of this proposition easily follows from the fact that $\mult$ is a many to one map and not one to many.

\begin{definition}[Homogeneous polynomials]
  A polynomial $P$ of degree $d$ is said to be homogeneous if it can be expressed as a linear combination of just the monomials of degree equal to $d$.
\end{definition}

\begin{definition}[Set-multilinear polynomials]
  A polynomial $P$ is said to be set-multilinear with respect to a set of variables $X$, under the partition $X=X_1\sqcup X_2\sqcup \ldots X_d$ if every monomial $m$ in the monomial support of $P$ is such that $\abs{\monsupp{m}\cap X_i}\leq 1$ for all $i\in [d]$.
\end{definition}

\begin{definition}[Multi-$r$-ic polynomials]
  A polynomial $P$ is said to be \mrc polynomial if the degree of the polynomial with respect to each of its variables is at most $r$.
\end{definition}

The following lemma (from \cite{GKKS}) is key to the asymptotic estimates required for the lower bound analyses. 
\begin{lemma}[Lemma 6, \cite{GKKS}]
  \label{lem:bin-gkks}
  Let $a(n), f(n), g(n) : \mathbb{Z}_{\geq 0} \rightarrow \mathbb{Z}_{\geq 0}$ be integer valued functions such that $(f+g) = o(a)$. Then, 
  \begin{align*}
    \ln \frac{(a+f)!}{(a-g)!} = (f+g)\ln a \pm O\left(\frac{(f+g)^2}{a}\right)
  \end{align*}
\end{lemma}

We shall now state a few lemmas that help us relate both the complexity measures introduced above.
\begin{lemma}[Observation 4.5 in \cite{FKS16}]
  \label{lem:sml-pd-eval-equiv}
  Let $X = X_1\sqcup\ldots\sqcup X_d$ and $\abs{X}=N$. Let $Y = X_1\sqcup \ldots\sqcup X_k$ for some $k\ll d$. Let $P$ be a homogeneous set multilinear polynomial of degree $d$ with respect to the partition $X_1\sqcup\ldots\sqcup X_d$. Let $m = Y^{\vece}$ be a set multilinear monomial\footnote{Here $\vece$ is a $\abs{Y}$-long vector that indicates the support of multilinear monomials. $Y^{\vece}$ is a shorthand representation of $y_1^{e_1}y_2^{e_2}\ldots y_{\abs{Y}}^{e_{\abs{Y}}}\,.$} of degree $k$ over $Y$. Then,
  \begin{align*}
    \frac{\partial^k P}{\partial Y^\vece} = P(\vece, Z).
  \end{align*}
\end{lemma}

\begin{corollary}[Similar to Corollary 4.6 in \cite{FKS16}]
  \label{cor:psspdml-sed}
  For a homogeneous and set multilinear polynomial $P(Y,Z)$ which is as defined as in \cref{lem:sml-pd-eval-equiv}, and for all values of parameters $k$ and $\ell$,
  \begin{align*}
    \psspd(P(Y,Z)) \leq \sed(P(Y,Z))\,.
  \end{align*}
\end{corollary}
\begin{proof}
  From the definition of the polynomial as defined in \cref{lem:sml-pd-eval-equiv}, it is easy to see that $\sigma_Y(\partial^{< k}P) = 0$. Further from \cref{lem:sml-pd-eval-equiv}, we know that
  \begin{align*}
    \sigma_Y\inparen{\partial_Y^{=k}P} &= \inbrace{P(\vece, Z)\mid \vece \in \zo^{\abs{Y}}_{= k} ~\text{indexes a set multilinear monomial over $Y$}}\\
    &\subseteq \inbrace{P(\vece, Z)\mid \vece \in \zo^{\abs{Y}}_{\leq k}}
  \end{align*}
  Multiplying both sides with the set $Z^{=\ell}$, we get the following.
  \begin{align*}
    Z^{=\ell}\cdot\sigma_Y\inparen{\partial_Y^{=k}P} &\subseteq Z^{=\ell}\cdot\inbrace{P(\vece, Z)\mid \vece \in \zo^{\abs{Y}}_{\leq k}}\,.
  \end{align*}
  Note that this inclusion continues to hold even after a multilinear projection.
  \begin{align*}
    \mult\inparen{Z^{=\ell}\cdot\sigma_Y\inparen{\partial_Y^{=k}P}} &\subseteq \mult\inparen{Z^{=\ell}\cdot\inbrace{P(\vece, Z)\mid \vece \in \zo^{\abs{Y}}_{\leq k}}}\,.
  \end{align*}
  Now taking the evaluation perspective of all the multilinear polynomials in the subspaces on both sides, we get that
  \begin{align*}
    \eval_{\zo}^{[Z]}\inparen{\mult\inparen{Z^{=\ell}\cdot\sigma_Y\inparen{\partial_Y^{=k}P}}} &\subseteq \eval_{\zo}^{[Z]}\inparen{\mult\inparen{Z^{=\ell}\cdot\inbrace{P(\vece, Z)\mid \vece \in \zo^{\abs{Y}}_{\leq k}}}}
  \end{align*}
  and thus
  \begin{align*}
    \dim\inparen{\eval_{\zo}^{[Z]}\inparen{\mult\inparen{Z^{=\ell}\cdot\sigma_Y\inparen{\partial_Y^{=k}P}}}} \leq \sed(P(Y,Z))\,.
  \end{align*}
  Putting this together with \cref{clm:eval-sync-mlequiv}, we get the following.
  \begin{align*}
    \psspd(P(Y,Z)) &= \dim\inparen{\F\text{-span}\inbrace{\mult\inparen{Z^{= \ell}\cdot \sigma_Y\inparen{\partial^{\leq k}_Y f}}}}\\
    &= \dim\inparen{\F\text{-span}\inbrace{\mult\inparen{Z^{= \ell}\cdot \sigma_Y\inparen{\partial^{=k}_Y f}}}}\\
		     &= \dim\inparen{\eval_{\zo}^{[Z]}\inparen{\mult\inparen{Z^{=\ell}\cdot\sigma_Y\inparen{\partial_Y^{=k}P}}}}\\
    &\leq \sed(P(Y,Z))\,.
  \end{align*}
\end{proof}

\begin{lemma}[Lemma 4.7 in \cite{FKS16}]
  Let $P(Y,Z)$ be a multi-$r$-ic polynomial. Then for every choice of parameters $k$ and $\ell$, we have
  \begin{align*}
    \inbrace{P(\vece, Z)\mid \vece \in \zo^{\abs{Y}}_{\leq k}} \subseteq \fspan\inbrace{\sigma_Y(\partial_Y^{\leq rk} P)}.
  \end{align*}
\end{lemma}

\begin{corollary}[Similar to Lemma 4.8 in \cite{FKS16}]
  \label{cor:sed-psspdr}
  For a multi-$r$-ic polynomial $P(Y,Z)$,
  \begin{align*}
    \sed(P(Y,Z)) \leq \psspdr(P(Y,Z)).
  \end{align*}
\end{corollary}
\begin{proof}
  \begin{align*}
    \inbrace{P(\vece, Z)\mid \vece \in \zo^{\abs{Y}}_{\leq k}} &\subseteq \fspan\inbrace{\sigma_Y(\partial_Y^{\leq rk} P)}
  \end{align*}
  Multiplying these polynomials on either sides by monomials in $Z^{=\ell}$, we get the following.
  \begin{align*}
    Z^{=\ell}\cdot\inbrace{P(\vece, Z)\mid \vece \in \zo^{\abs{Y}}_{\leq k}} &\subseteq \fspan\inbrace{Z^{=\ell}\cdot\sigma_Y(\partial_Y^{\leq rk} P)}.
  \end{align*}
  Note that this inclusion continues to hold under multilinear projections.
  \begin{align*}
    \mult\inparen{Z^{=\ell}\cdot\inbrace{P(\vece, Z)\mid \vece \in \zo^{\abs{Y}}_{\leq k}}} &\subseteq \fspan\inbrace{\mult\inparen{Z^{=\ell}\cdot\sigma_Y(\partial_Y^{\leq rk} P)}}.
  \end{align*}
  Putting this together with \cref{clm:eval-sync-mlequiv}, we get the following.
  \begin{align*}
    \sed(P(Y,Z)) &= \dim\inparen{\eval^{[Z]}_{\zo}\inbrace{\mult\inparen{Z^{=\ell}\cdot\inbrace{P(\vece, Z)\mid \vece \in \zo^{\abs{Y}}_{\leq k}}}}}\\
		  &= \dim\inparen{\fspan\inbrace{\mult\inparen{Z^{=\ell}\cdot\inbrace{P(\vece, Z)\mid \vece \in \zo^{\abs{Y}}_{\leq k}}}}}\\
		  &\leq \dim\inparen{\fspan\inbrace{\mult\inparen{Z^{=\ell}\cdot\sigma_Y(\partial_Y^{\leq rk} P)}}}\\
    &= \psspdr(P(Y,Z))\,.
  \end{align*}
\end{proof}

\subsection*{Complexity measure for the $\SWSP$ circuits of low bottom support}
\begin{lemma}\label{lem:cktub-pow}
  Let $m,k, \ell$ and $t$ be positive integers such that $\ell +kt < \frac{m}{2}$. Let $Y$ and $Z$ be disjoint sets of variables such that $\abs{Z}=m$. Let $C(Y,Z)$ be a depth four $\SWSP$ circuit of bottom support at most $t$ with respect to variables from $Z$, and size $s$. Then, $\psspd(C)$ is at most $s\cdot(k+1)\cdot{m\choose \ell+kt}\cdot(\ell+kt)$.
\end{lemma}
\begin{proof}
  Let $C(Y,Z)$ be equal to the sum $T_1(Y,Z)+\ldots+T_s(Y,Z)$ where $T_i(Y,Z)=\inparen{Q_i(Y,Z)}^{e_i}$ ($i\in [s]$, $e\in \mathbb{Q}$ and $Q_i\in \F[Y\sqcup Z]$ is a polynomial each of whose monomials are supported on at most $t$ many variables from $Z$). It is easy to verify that the measure of Projected Skew Partial derivatives is sub-additive and thus we get that
  \begin{equation}
    \psspd(C(Y,Z)) \leq \sum_{i\in[s]}\psspd(T_i(Y,Z))\,.\label{eqn:subadd-pow}  
  \end{equation}
  Let $T(Y,Z)$ be an arbitrary term in $\inbrace{T_1(Y,Z),\ldots, T_s(Y,Z)}$ such that $T=\inparen{Q(Y,Z)}^e$ for some $Q(Y,Z)$ all of whose monomials are supported on at most $t$ many variables from $Z$, and $e\in\mathbb{Q}$.

  \paragraph*{Case when $e\geq k$:} We shall prove by induction on $k$ that for any monomial $m\in\F[Y]$ of degree $k$, 
  \begin{align*}
    \partial^{k}_m(T(Y,Z)) \in \fspan\inbrace{\inparen{Q(Y,Z)}^{e-k}\cdot\inbrace{Z_{\inbrace{\leq kt}}}\cdot\F[Y]}.
  \end{align*}
  Base case when $k=0$ is trivial as $T$ is already in the required form $(Q(Y,Z))^e\cdot Z_{\inbrace{=0}}\cdot 1$. Now assume the induction hypothesis for all $\partial^{k'}_{m'}$ ($k' \leq k-1$). Let $m'=y_{i_1}\ldots y_{i_{k-1}}$ be a monomial in $\F[Y]$ and $\partial^{k-1}_{m'}T$ be expressed as $(Q(Y,Z))^{e-(k-1)}\cdot g(Z) \cdot h(Y)$ where $g(Z)$ is a polynomial in $\F[Z]$ each of whose monomials are supported on at most $(k-1)t$ many variables, and $h(Y)$ is some arbitrary polynomial in $\F[Y]$. Further deriving $\partial^{k-1}_{m'}T$ with $y_{i_k}$, we get the following.
  \begin{align*}
    \frac{\partial\inparen{\partial^{k-1}_{m'}T}}{\partial y_{i_k}} &= (e-k+1)\cdot (Q(Y,Z))^{e-(k-1)-1}\cdot \frac{\partial Q(Y,Z)}{\partial y_{i_k}}\cdot g(Z) \cdot h(Y)\\ &\quad \quad +(Q(Y,Z))^{e-(k-1)-1}\cdot Q(Y,Z) \cdot g(Z)\cdot \frac{\partial h(Y)}{\partial y_{i_k}}\\
								 &= (Q(Y,Z))^{e-k}\cdot\inparen{(e-k+1)\cdot\frac{\partial Q(Y,Z)}{\partial y_{i_k}}\cdot h(Y) + Q(Y,Z)\cdot\frac{\partial h(Y)}{\partial y_{i_k}}}\cdot g(Z)\\
								 &\in \inbrace{(Q(Y,Z))^{e-k}\cdot g(Z) \cdot Z_{\inbrace{\leq t}}\cdot \F[Y]}\\
    &\subseteq \fspan\inbrace{(Q(Y,Z))^{e-k}\cdot\inbrace{Z_{\inbrace{\leq kt}}}\cdot\F[Y]}\,.
  \end{align*}
  The inclusion in the third line of the math block above follows from the fact that both the polynomial $Q(Y,Z)$ and its derivative $\frac{\partial Q(Y,Z)}{\partial y_{i_k}}$ can be expressed as $(\F[Y])$-linear combinations of monomials in $Z_{\inbrace{\leq t}}$, and the inclusion in the last line follows from the fact that $g(Z)\in \F[Z]$ is a polynomial each of whose monomials are supported on at most $(k-1)t$ many variables from $Z$. Thus,
  \begin{align*}
    \partial^{\leq k} T(Y,Z) &\subseteq \fspan\inbrace{\inbrace{(Q(Y,Z))^a\mid a\in[e-k,e]}\cdot\inbrace{Z_{\inbrace{\leq kt}}}\cdot\F[Y]}\,.
  \end{align*}
  Applying the projection $\sigma_Y$, the shift $Z^{=\ell}$, and multilinear projection $\mult$ on both sides, we get that 
  \begin{align*}
    \fspan\inbrace{\mult\inparen{Z^{=\ell}\cdot \sigma_Y(\partial^{\leq k}T(Y,Z))}} &\subseteq \fspan\inbrace{\mult\inparen{\inbrace{(\sigma_Y(Q(Y,Z)))^a\mid a\in[e-k, e]}\cdot\inbrace{Z_{\inbrace{\leq \ell+kt}}}}}\\
										    &\subseteq \fspan\inbrace{\mult\inparen{\inbrace{(\sigma_Y(Q(Y,Z)))^a\mid a\in[e-k,e]}}\cdot Z^{\leq \ell+kt}_{\ml}}.
  \end{align*}
  The last inclusion follows from Item 2 of \cref{prop:mult}. This implies that
  \begin{align*}
    \dim\inparen{\fspan\inbrace{\mult\inparen{Z^{=\ell}\cdot \sigma_Y(\partial^{\leq k}T(Y,Z))}}} &\leq \dim\inparen{\mult\inparen{\inbrace{(\sigma_Y(Q(Y,Z)))^a\mid a\in[e-k,e]}}}\\
    &\quad \quad \cdot \dim(Z^{\leq \ell+kt}_{\ml})\,.
  \end{align*}
  Here $\dim\inparen{\mult\inparen{\inbrace{(\sigma_Y(Q(Y,Z)))^a\mid a\in[e-k,e]}}}$ is at most $(k+1)$, and $\dim(Z^{\leq \ell+kt}_{\ml})$ is at most ${m\choose \ell+kt}\cdot (\ell+kt)$ when $\ell+kt\leq \frac{m}{2}$.
  Thus, $\psspd(T(Y,Z))$ is at most $(k+1)\cdot {m\choose \ell+kt}\cdot (\ell+kt)$ when $\ell+kt\leq \frac{m}{2}$. 
  \paragraph*{Case when $e<k$:} It is easy to see that $Q^e$ and its partial derivatives of any order with respect to variables from $Y$ can be expressed as a $(\F[Y])$-linear combinations of monomials in $Z_{\inbrace{\leq kt}}$. Thus,
  \begin{align*}
    \fspan\inbrace{\mult\inparen{Z^{=\ell}\cdot\sigma_Y(\partial^{\leq k} T(Y,Z)}} \subseteq \fspan\inbrace{\mult\inparen{Z_{\inbrace{\leq \ell+kt}}}} \subseteq \fspan\inbrace{Z^{\leq \ell+kt}_{\ml}}\,.
  \end{align*}
  Thus, $\psspd(T(Y,Z))$ in this case is at most $(\ell+kt)\cdot {m\choose \ell+kt}$ when $\ell+kt <\frac{m}{2}$.

  Putting both of these cases together with \cref{eqn:subadd-pow} and the fact that $\ell+kt\leq \frac{m}{2}$, we get that
  \begin{align*}
    \psspd(C(Y,Z)) \leq \sum_{i\in [s]}\psspd(T_i(Y,Z))&\leq s\cdot\max_{i\in [s]}\inbrace{\psspd(T_i(Y,Z))}\\ &\leq s\cdot (k+1)\cdot {m\choose \ell+kt}\cdot (\ell+kt).
  \end{align*}
  This completes the proof.
\end{proof}

\section{Hard Polynomial and Restrictions}
In this section we recall the definition of the polynomial family and the set of deterministic and random restrictions imposed on the polynomial family, from \cite{ChiExp20}.

\subsection{Polynomial Family: Iterated Matrix Multiplication polynomial}
Let $X^{(1)}, X^{(2)}, \ldots, X^{(d)}$ be $d$ generic $n\times n$ matrices defined over disjoint set of variables. For any $k\in[d]$, let $x_{i,j}^{(k)}$ be the variable in the matrix $X^{(k)}$ indexed by $(i,j)\in[n]\times[n]$. The Iterated Matrix Multiplication polynomial, denoted by the family $\{\IMM_{n,d}\}$, is defined as follows.  
  \begin{align*}
    \IMM_{n,d}(X) = \sum_{i_1, i_2, \ldots, i_{d-1} \in [n]}x_{1,i_1}^{(1)}x_{i_1,i_2}^{(2)}\dots x_{i_{(d-2)},i_{(d-1)}}^{(d-1)}x_{i_{(d-1)},1}^{(d)}.
  \end{align*}

\subsection{Deterministic and Random Restrictions}

Let $k$ and $\alpha$ be a parameters such that $d = (2\alpha+3)\cdot k$. Let the $d$ matrices be divided into $k$ contiguous blocks of matrices $B_1, B_2, \dots, B_{k}$ such that each block $B_i$ contains $2\alpha+3$ matrices. By suitable renaming, let us assume that each block $B_i$ contains the following matrices.
\begin{align*}
X^{(i,L,\alpha+1)}, \cdots, X^{(i,L,2)},X^{(i,L,1)}, X^{(i)},X^{(i,R,1)}, X^{(i,R,2)}, \cdots, X^{(i,R,\alpha+1)}.
\end{align*}

Let us first consider the following set of restrictions, first deterministic and then randomized. 

\subsubsection*{Deterministic Restrictions}
Let $V_0:X\mapsto Y_0\sqcup Z_0\sqcup \inbrace{0,1}$ be a deterministic restriction of the variables $X$ in to disjoint variable sets $Y_0$, $Z_0$, and $\zo$ as follows. For all $i\in[k]$, 
\begin{itemize}
\item The variables in matrix in $X^{(i)}$ are each set to a distinct $Y_0$ variable. Henceforth, we shall refer to this as $Y^{(i)}$ matrix.
\item The entries of the first row of matrix $X^{(i,L,\alpha+1)}$ are all set to $1$ and the rest of the matrix to $0$.
\item The entries of the first column of matrix $X^{(i,R,\alpha+1)}$ are all set to $1$ and the rest of the matrix to $0$.
\item The rest of the variables are all set to distinct $Z_0$ variables. Henceforth, for all $b\in\inbrace{L,R}$ and $j\in[\alpha]$, we shall refer to the matrix $X^{(i,b,j)}$ as $Z^{(i,b,j)}$ matrix.
\end{itemize}

\subsubsection*{Random Restrictions}
Let $\eta$ and $\varepsilon'$ be two fixed constants in $(0,1)$. Let $V_1:Y_0\sqcup Z_0\mapsto Y\sqcup Z\sqcup \inbrace{0,1}$ be a random restriction of the variables $Y_0\sqcup Z_0$ as follows.
\begin{itemize}
\item Matrix $Z^{(i,L,1)}$: For every column, pick $n^{\eta}$ distinct elements uniformly at random and keep these elements alive. Set the other entries in this matrix to zero.
\item Matrix $Z^{(i,R,1)}$: For every row, pick $n^{\eta}$ distinct elements uniformly at random and keep these elements alive. Set the other entries in this matrix to zero.
\item Matrices $Z^{(i,L,j)}$ for all $j\in[2, \alpha-\varepsilon'\log{n}]$: For every column, pick $2$ distinct elements uniformly at random and set all the other entries to zero. 
\item Matrices $Z^{(i,R,j)}$ for all $j\in[2, \alpha-\varepsilon'\log{n}]$: For every row, pick $2$ distinct elements uniformly at random and set all the other entries to zero.
  \item Matrices $Z^{(i,L,j)}$ for all $j>\alpha-\varepsilon'\log{n}$: For every column, pick $1$ element uniformly at random and set the other elements in that row to zero.
\item Matrices $Z^{(i,R,j)}$ for all $j>\alpha-\varepsilon'\log{n}$: For every row, pick $1$ element uniformly at random and set the other elements in that row to zero.
\end{itemize}

Let $D$ be the distribution of all the restrictions $V:X\mapsto Y\sqcup Z\sqcup\zo$ such that $V = V_1\circ V_0$ where $V_0$ and $V_1$ are deterministic and random restrictions respectively, as described above. Let $m$ be used to denote the number of $Z$ variables left after the restriction and $m = 2kn(n^{\eta}+2(\alpha-\varepsilon'\log{n}-1)+\varepsilon'\log{n}) = O(n^{1+\eta}k)$ when $\alpha \leq O(n^\eta)$.

\subsection*{Effect of Restrictions on $\IMM_{n,d}$}
Let $g^{(i,L)}_{1,a}(Z)$ be the $(1,a)$th entry in product of matrices $\prod_{j = 0}^{\alpha}X^{(i,L,\alpha+1-j)}|_V$. Let $g^{(i,R)}_{b,1}(Z)$ be the $(b,1)$th entry in product of matrices $\prod_{j = 1}^{\alpha+1}X^{(i,R,j)}|_V$. Let $g^{(i)}$ the $(1,1)$th entry in the product of all the matrices in the block $B_i$. Then we can express $g^{(i)}$ as follows.
\begin{align*}
  g^{(i)}(Y,Z) = \sum_{a,b\in[n]}g^{(i,L)}_{1,a}(Z)\cdot y^{(i)}_{a,b}\cdot g^{(i,R)}_{b,1}(Z).
\end{align*}
Let $P|_V(Y,Z)$ obtained by restricting $\IMM_{n,d}(X)$ with the restriction $V \leftarrow D$. Thus,
\begin{align*}
  P|_V (Y,Z) = \prod_{i=1}^{k}g^{(i)}(Y,Z)\,.
\end{align*}


To summarize, for some parameters $\alpha,k,\eta$ and $m$, $P|_V$ is polynomial in $\F[Y\sqcup Z]$ such that its degree is $d=(2\alpha+3)\cdot k$, and has $m=O(n^{1+\eta}k)$ many $Z$ variables. Here the definition of the polynomial $P|_V$ is heavily dependent on $V\leftarrow D$ and the choice of parameters $\alpha, k, \varepsilon'$ and $\eta$.

\paragraph*{Effect on random restrictions:}
\begin{lemma}[Lemma 8, \cite{ChiExp20}]
  \label{lem:ckt-randrest}
  Let $t$ be a parameter. Let $C$ be any depth four circuit of size at most $s\leq n^{\frac{t}{2}}$ that computes $\IMM_{n,d}$. Then with a probability of at least $1-o(1)$, over $V\leftarrow D$ (where $V:X\mapsto Y\sqcup Z\sqcup\zo$), $C|_{V}$ is a depth four circuit of bottom support at most $t$ in $Z$ variables that computes the polynomial $P|_V(Y,Z)$.
\end{lemma}

\subsection{Complexity of $P|_V$}
\subsubsection*{Choice of parameters}
We borrow the setting of the parameters involved directly from \cite{ChiExp20}\footnote{In an attempt to have a clean up the notation in comparison to \cite{ChiExp20}, we make the following notational changes -- the parameter $\alpha$ here corresponds to $k'$ in \cite{ChiExp20}, the parameter $k$ here corresponds to $r'$ in \cite{ChiExp20}. Further the parameter $k=d-3r'=2k'r'$ in \cite{ChiExp20} translates to $2\alpha k$ here. The rest of the parameters $\varepsilon, \varepsilon', \eta$ and $\tau$ are the same in both the papers.}.
\begin{multicols}{2}
  \begin{itemize}
  \item $\varepsilon' = 0.34$,
  \item $\eta = 0.05$,
  \item $\varepsilon= \varepsilon'-\eta = 0.29$,
  \item $\tau=0.08$,
  \item $\omega(\log{n})\leq d \leq n^{0.01}$,
  \item $d = (2\alpha+3)k$,
  \item $m=\Theta(n^{1+\eta}k) = \Theta(n^{1.05}k)$,
  \item $\ell = \frac{m}{2}(1-\Gamma)$,
  \item $(1+\Gamma)^\alpha = 2n^\varepsilon$ such that $\Gamma = O_{\varepsilon}\inparen{\frac{\ln n}{\alpha}}$,

  \end{itemize}
\end{multicols}

We shall now recall the following from \cite{ChiExp20}.
\begin{theorem}[Discussion above Theorem 17, \cite{ChiExp20}]
  \label{thm:psspd-chiexp}
  Let $n$ be a large enough integer. Let $m, d, \ell, \alpha, k, \varepsilon$ and $\tau$ be as described above. 
\begin{align*}
 \psspd(P|_V) \geq \frac{\inparen{\frac{m}{m-\ell}}^{2\alpha k}\cdot {m-2\alpha k\choose \ell}}{2^{O(k)}\cdot\inparen{\frac{\ell}{m-\ell}}^{2\alpha k(1-\tau)}}.
\end{align*}
\end{theorem}

Note that for a $N$-variate polynomial $P(X,Y)$, the measure in \cite{ChiExp20} was defined to be equal to $\dim\inparen{\F\text{-span}\inbrace{\mult_0\inparen{Z^{= \ell}\cdot \sigma_Y\inparen{\partial^{= k}_Y P}}}}$ where $\mult_0(P) = P \mod \inbrace{x_i^2 \mid i\in [N]}$ compared to the measure here which is equal to $\dim\inparen{\F\text{-span}\inbrace{\mult\inparen{Z^{= \ell}\cdot \sigma_Y\inparen{\partial^{\leq k}_Y P}}}}$ where $\mult(P) = P \mod \inbrace{x_i^2-x_i \mid i\in [N]}$. This change of definition would not affect the bound as the lower bound in \cite{ChiExp20} counts the leading monomials of support size and degree both equal to $d-k+\ell$, and $\sigma_Y(\partial^{< k} P|_V) = \emptyset$ for the polynomial $P|_V$ described above.

\section{Functional Lower Bounds against restricted $\SWSP$ Circuits}

As mentioned in the proof overview, we first prove a lower bound against bounded bottom support depth four circuits and then escalate this lower bound to circuits without the restriction on bottom support.

\begin{lemma}
  \label{lem:fn-pow-d4-bottom-PV}
  Let $n$ and $d$ be large integers such that $\omega(\log^2n)\leq d \leq n^{0.01}$. Let $\alpha, k, r$ and $t$ be parameters such that $d=(2\alpha+3)k$ and $r\leq \frac{\alpha}{200t}$. Any depth four $\SWSP$ circuit of bounded individual degree $r$ and bounded bottom fan-in at most $t$, computing a function equivalent to $P|_V(X_V)$ (for $V\leftarrow D$) on $\zo^{\abs{X_V}}$, must have size at least $n^{\Omega(k)}$.
\end{lemma}
\begin{proof}
  Let $C(Y,Z)$ be a $\SWSP$ circuit of bounded individual degree $r$, bottom fan-in at most $t$ and size $s$. Since the polynomial computed at the root of circuit $C(Y,Z)$ is functionally equivalent to $P|_V(Y,Z)$, we get that
  \begin{align*}
    \sed(P|_V(Y,Z)) = \sed(C(Y,Z)).
  \end{align*}
  Further, from \cref{cor:psspdml-sed} and \cref{cor:sed-psspdr}, the above equation can be extended to the following inequality.
  \begin{equation}
    \label{eqn:chain-ineq-D-pow}
    \psspd(P|_V(Y,Z)) \leq \sed(P|_V(Y,Z)) = \sed(C(Y,Z)) \leq \psspdr(C(Y,Z)).
  \end{equation}
  From \cref{thm:psspd-chiexp}, we have that
  \begin{equation}\label{eqn:fn-lb-D-pow}
    \psspd(P|_V(Y,Z)) \geq \frac{\inparen{\frac{m}{m-\ell}}^{2\alpha k}\cdot {m-2\alpha k\choose \ell}}{2^{O(k)}\cdot\inparen{\frac{\ell}{m-\ell}}^{2\alpha k(1-\tau)}}
  \end{equation}
  and from \cref{lem:cktub-pow}, we have that
  \begin{equation}\label{eqn:fn-ub-D-pow}
    \psspdr(C(Y,Z)) \leq s\cdot(kr+1)\cdot{m\choose \ell+krt}\cdot{(\ell+krt)}.
  \end{equation}

  Putting \cref{eqn:chain-ineq-D-pow}, \cref{eqn:fn-lb-D-pow} and \cref{eqn:fn-ub-D-pow} together, we get the following.
  \begin{align*}
    \frac{\inparen{\frac{m}{m-\ell}}^{2\alpha k}\cdot {m-2\alpha k\choose \ell}}{2^{O(k)}\cdot\inparen{\frac{\ell}{m-\ell}}^{2\alpha k(1-\tau)}}\leq s\cdot (kr+1)\cdot {m\choose \ell+krt}\cdot{(\ell+krt)}.
  \end{align*}
  Thus,
  \begin{align*}
    s &\geq \frac{\inparen{\frac{m}{m-\ell}}^{2\alpha k}\cdot {m-2\alpha k\choose \ell}}{2^{O(k)}\cdot\inparen{\frac{\ell}{m-\ell}}^{2\alpha k(1-\tau)}}\times\frac{1}{(kr+1)\cdot{m\choose \ell+krt}\cdot{(\ell+krt)}}\\
      &= \frac{\inparen{\frac{m}{m-\ell}}^{2\alpha k}}{2^{O(k)}}\cdot\frac{(m-2\alpha k)!}{\ell!\cdot (m-\ell-2\alpha k)!}\cdot\frac{(\ell+krt)!(m-\ell-krt)!}{m!}\cdot\inparen{\frac{m-\ell}{\ell}}^{2\alpha k(1-\tau)}\\
      &= \frac{\inparen{\frac{m}{m-\ell}}^{2\alpha k}}{2^{O(k)}}\cdot\frac{(m-2\alpha k)!}{m!}\cdot\frac{(m-\ell)!}{(m-\ell-2\alpha k)!}\cdot\frac{(\ell+krt)!}{\ell!}\cdot\frac{(m-\ell-krt)!}{(m-\ell)!}\cdot\inparen{\frac{m-\ell}{\ell}}^{2\alpha k(1-\tau)}\\
      &\approx \frac{\inparen{\frac{m}{m-\ell}}^{2\alpha k}}{2^{O(k)}}\cdot\inparen{\frac{m-\ell}{m}}^{2\alpha k}\cdot\inparen{\frac{\ell}{m-\ell}}^{krt}\cdot\inparen{\frac{m-\ell}{\ell}}^{2\alpha k(1-\tau)}\\
      &= \inparen{\frac{m-\ell}{\ell}}^{2\alpha k(1-\tau) - krt}\cdot\frac{1}{2^{O(k)}}\\
      &= \inparen{\frac{1+\Gamma}{1-\Gamma}}^{2\alpha k(1-\tau) - krt}\cdot\frac{1}{2^{O(k)}}\\
      &\geq \inparen{(1+\Gamma)^2}^{2\alpha k(1-\tau) - krt}\cdot\frac{1}{2^{O(k)}}\\
      &= \inparen{(1+\Gamma)^\alpha}^{4k(1-\tau) - \frac{2krt}{\alpha}}\cdot\frac{1}{2^{O(k)}}\\
      &\approx \inparen{2n^\varepsilon}^{4k(1-\tau) - \frac{2krt}{\alpha}}\cdot\frac{1}{2^{O(k)}}\\
      &= {\inparen{{\Theta(1)}\cdot n^{4\varepsilon(1-\tau) - 10^{-2}}}^{k}}\\
      &\geq {n^{1.05k}}.
  \end{align*}
  In the above math block, in line 2 we absorb $(kr+1)\cdot(\ell+krt)$ into $2^{O(k)}$, in line 4 we use \cref{lem:bin-gkks} to get the approximations, in line 6 we use the fact that $m = \frac{\ell}{2}(1-\Gamma)$, in line -3 we use the fact that $(1+\Gamma)^\alpha \approx 2n^{\varepsilon}$, in line -2, we use the fact that $r$ is at most $\frac{\alpha}{200t}$, and in the last line we use the fact that $\varepsilon=0.29$ and $\tau=0.08$.
\end{proof}

\subsection*{Proof of \cref{thm:fn-pow-d4}}
For a large integer $n$, let $d$ be such that $\Omega(\log^2n) \leq d \leq n^{0.01}$. Let $t$ be a parameter that we shall soon fix. Let $C$ be a $\SWSP$ circuit of bounded individual degree at most $r$, and size $s \leq n^{\frac{t}{2}}$ that computes a polynomial $Q(X)$ that is functionally equivalent to $\IMM_{n,d}(X)$ (over $\zo^{n^2d}$). Let $\alpha$ and $k$ be parameters such that $d = (2\alpha+3)k$.  Recall that a restriction $V\leftarrow D$ fixes a subset of variables to values in $\zo$ and maps the rest to distinct $Y$ and $Z$ variables. For any such restriction $V\leftarrow D$, let $X_V = Y\sqcup Z$ be the set of variables in $X$ that are not set to values in $\zo$ by $V$. From \cref{lem:ckt-randrest} we know that with a probability of at least $1-o(1)$, the circuit $C_V$ obtained by applying the restriction $V$ to $C$ is a $\SWSP$ circuit of bounded individual degree at most $r$, size $s$ and bottom support at most $t$. Let $Q_V$ be the polynomial computed by $C_V$, over $X_V$ variables. We shall now show that $Q_V$ is functionally equivalent to $P|_V$ over $\zo^{\abs{X_V}}$.

Let the set $S_V\subset \zo^{n^2d}$ be the subset of points such that for all $\veca \in S_V$, if $x_i\in X\setminus X_V$ and $V$ sets $x_i$ to $b\in\zo$, then the value at the $i$'th location of $\veca$, $\veca_i = b$. Since $Q(X)$ and $\IMM_{n,d}(X)$ are functionally equivalent over all of $\zo^{n^2d}$, they are functionally equivalent over $S_V$ as well. Thus, $Q_V(\veca|_{X_V}) = Q(\veca) = \IMM(\veca) = P|_V(\veca|_{X_V})$ for all $\veca \in S_V$. Here $\veca|_{X_V}\in \zo^{\abs{X_V}}$ corresponds to projection of $\veca\in\zo^{n^2d}$ to locations corresponding to the variables in $X_V$.

This implies that $Q_V(X_V)$ and $P|_V(X_V)$ are functionally equivalent over $\zo^{\abs{X_V}}$ and thus, there is a $\SWSP$ circuit of bounded individual degree at most $r$, size $s\leq n^{\frac{t}{2}}$ and bottom support at most $t$ that functionally computes $P|_V(Y,Z)$. On the other hand if $r$ is at most $\frac{\alpha}{200t}$ then from \cref{lem:fn-pow-d4-bottom-PV} we know that any $\SWSP$ circuit of bounded individual degree at most $r$ and bottom support at most $t$ that functionally computes $P|_V$ must have size $n^{\Omega\inparen{k}}$. Putting these together by fixing the value of $t$ to $3k$ we get that $s$ must at least be $n^{\Omega\inparen{k}}$. Since $r$ is at most $\frac{\alpha}{200t}$, under this substitution of $t$, this value computes to $\frac{1}{200\cdot 3k}\cdot\inparen{\frac{d}{2k} - \frac{3}{2}} = \frac{d}{1200k^2} - \frac{1}{400k}$.
\qed



\section{Syntactic circuit lower bounds against $\SWSP$ circuits}
We shall again prove a lower bound against circuits of low bottom support and then escalate this bound to circuits without any restriction on bottom support.
\begin{lemma}
  \label{lem:syn-d4-bottom-PV}
  Let $n$ and $d$ be large integers such that $\omega(\log^2n)\leq d \leq n^{0.01}$. Let $\alpha$ and $k$ be parameters such that $d=(2\alpha+3)k$. Any depth four $\SWSP$ circuit of bounded bottom fan-in at most $t= \frac{\alpha}{200}$, syntactically computing $P|_V(X_V)$ (for $V\leftarrow D$) must have size at least $n^{1.05k}$.
\end{lemma}
\begin{proof}
  Let $C(Y,Z)$ be a $\SWSP$ circuit of bounded individual degree $r$, bottom fan-in at most $t$ and size $s$. Since the polynomial computed at the root of circuit $C$ is functionally equivalent to $P|_V(Y,Z)$, we get that
  \begin{equation}
    \label{eqn:equiv-pow-psspd}
    \psspd(P|_V(Y,Z))  \leq \psspd(C(Y,Z)).
  \end{equation}
  From \cref{thm:psspd-chiexp}, we have
  \begin{equation}\label{eqn:syn-lb-D-pow}
    \psspd(P|_V(Y,Z)) \geq \frac{\inparen{\frac{m}{m-\ell}}^{2\alpha k}\cdot {m-2\alpha k\choose \ell}}{2^{O(k)}\cdot\inparen{\frac{\ell}{m-\ell}}^{2\alpha k(1-\tau)}}
  \end{equation}
  and from \cref{lem:cktub-pow}, we have that
  \begin{equation}\label{eqn:syn-ub-D-pow}
    \psspdr(C(Y,Z)) \leq s\cdot(k+1)\cdot{m\choose \ell+kt}\cdot{(\ell+kt)}.
  \end{equation}

  Putting \cref{eqn:equiv-pow-psspd}, \cref{eqn:syn-lb-D-pow} and \cref{eqn:syn-ub-D-pow} together, we get the following.
    \begin{align*}
    s\cdot (k+1)\cdot {m\choose \ell+kt}\cdot{(\ell+kt)} \geq \frac{\inparen{\frac{m}{m-\ell}}^{2\alpha k}\cdot {m-2\alpha k\choose \ell}}{2^{O(k)}\cdot\inparen{\frac{\ell}{m-\ell}}^{2\alpha k(1-\tau)}}.
  \end{align*}
  Thus,
  \begin{align*}
    s &\geq \frac{\inparen{\frac{m}{m-\ell}}^{2\alpha k}\cdot {m-2\alpha k\choose \ell}}{2^{O(k)}\cdot\inparen{\frac{\ell}{m-\ell}}^{2\alpha k(1-\tau)}}\times\frac{1}{(k+1)\cdot{m\choose \ell+kt}\cdot{(\ell+kt)}}\\
      &= \frac{\inparen{\frac{m}{m-\ell}}^{2\alpha k}}{2^{O(k)}}\cdot\frac{(m-2\alpha k)!}{\ell!\cdot (m-\ell-2\alpha k)!}\cdot\frac{(\ell+kt)!(m-\ell-kt)!}{m!}\cdot\inparen{\frac{m-\ell}{\ell}}^{2\alpha k(1-\tau)}\\
      &= \frac{\inparen{\frac{m}{m-\ell}}^{2\alpha k}}{2^{O(k)}}\cdot\frac{(m-2\alpha k)!}{m!}\cdot\frac{(m-\ell)!}{(m-\ell-2\alpha k)!}\cdot\frac{(\ell+kt)!}{\ell!}\cdot\frac{(m-\ell-kt)!}{(m-\ell)!}\cdot\inparen{\frac{m-\ell}{\ell}}^{2\alpha k(1-\tau)}\\
      &\approx \frac{\inparen{\frac{m}{m-\ell}}^{2\alpha k}}{2^{O(k)}}\cdot\inparen{\frac{m-\ell}{m}}^{2\alpha k}\cdot\inparen{\frac{\ell}{m-\ell}}^{kt}\cdot\inparen{\frac{m-\ell}{\ell}}^{2\alpha k(1-\tau)}\\
      &= \inparen{\frac{m-\ell}{\ell}}^{2\alpha k(1-\tau) - kt}\cdot\frac{1}{2^{O(k)}}\\
      &= \inparen{\frac{1+\Gamma}{1-\Gamma}}^{2\alpha k(1-\tau) - kt}\cdot\frac{1}{2^{O(k)}}\\
      &\geq \inparen{(1+\Gamma)^2}^{2\alpha k(1-\tau) - kt}\cdot\frac{1}{2^{O(k)}}\\
      &= \inparen{(1+\Gamma)^\alpha}^{4k(1-\tau) - \frac{2kt}{\alpha}}\cdot\frac{1}{2^{O(k)}}\\
      &\approx \inparen{2n^\varepsilon}^{4k(1-\tau) - \frac{2kt}{\alpha}}\cdot\frac{1}{2^{O(k)}}\\
      &= {\inparen{{\Theta(1)}\cdot n^{4\varepsilon(1-\tau) - 10^{-2}}}^{k}}\\
      &\geq {n^{1.05k}}.
  \end{align*}
  In the above math block, in line 2 we absorb $(k+1)\cdot(\ell+kt)$ into $2^{O(k)}$, in line 4 we use \cref{lem:bin-gkks} to get the approximations, in line 6 we use the fact that $m = \frac{\ell}{2}(1-\Gamma)$, in line -3 we use the fact that $(1+\Gamma)^\alpha \approx 2n^{\varepsilon}$, in line -2 we use the fact that $t$ is at most $\frac{\alpha}{200}$, and in the last line we use the fact that $\varepsilon=0.29$ and $\tau=0.08$.
\end{proof}

\subsection*{Proof of \cref{thm:syn-pow-d4}}
  Let $t$ be a parameter such that $t\geq 3k$ and $t\leq \frac{\alpha}{200}$. Let $\alpha$ and $k$ be such that $d=(2\alpha + 3)\cdot k$. Let $C$ be a $\SWSP$ circuit of size at most $n^{\frac{t}{2}}$ computing the $\IMM_{n,d}$ polynomial. From \cref{lem:ckt-randrest}, we get that with a probability of at least $(1-o(1))$ over $V\leftarrow D$, $C|_V$ is a $\SWSP$ circuit of bottom support at most $t$. Note that $C|_V$ is of size at most $n^{\frac{t}{2}}$. From \cref{lem:syn-d4-bottom-PV}, $C|_V$ must have size at least $n^{1.05k}$. From our choice of parameters, $1.05k$ is at most $\frac{t}{2}$. We choose the parameters $\alpha$ and $k$ to be in the order of $\Theta(\sqrt{d})$ such that $\alpha \geq 600k$. Thus, any $\SWSP$ circuit computing $\IMM_{n,d}$ must have size at least $n^{1.05k} = n^{\Omega(\sqrt{d})}$. 
\qed

\section{Functional lower bounds against restricted $\SPSP$ circuits}
Analogous to \cref{lem:cktub-pow}, we can also prove a bound on $\psspd(C)$ where $C$ is a $\SPSP$ circuit of bounded formal degree and bounded bottom support.

\begin{lemma}\label{lem:cktub}
  Let $n,k, r, \ell$ and $t$ be positive integers such that $\ell +kt < \frac{m}{2}$. Let $C(Y,Z)$ be a depth four circuit of formal degree at most $d $, bottom support at most $t$ with respect to $Z$ variables, and size $s$. Then, $\psspd(C)$ is at most $s\cdot{\frac{2d }{t}+1\choose k}\cdot{m\choose \ell+kt}\cdot(\ell+kt)$.
\end{lemma}

We shall again prove a lower bound on circuits of bounded bottom support and then escalate it to the model of interest.

\subsection*{Proof of \cref{thm:fn-prod-d4}}
Let $\alpha$ and $k$ be parameters such that $d = (2\alpha+3)k$. Let $t$ be a parameter that we shall soon fix so that it satisfies the criteria that $t\geq 0.1k$ and $r \leq \frac{\alpha}{200t}$. For a large integer $n$, let $d$ be such that $\Omega(\log^2n) \leq d \leq n^{0.01}$. Let $C$ be a $\SPSP$ circuit of bounded formal degree $d $, bounded individual degree at most $r$, and size $s \leq n^{\frac{t}{2}}$ that computes a polynomial $Q(X)$ that is functionally equivalent to $\IMM_{n,d}(X)$ (over $\zo^{n^2d}$).   

From \cref{lem:ckt-randrest} we know that with a probability of at least $1-o(1)$, the circuit $C_V$ obtained by applying the restriction $V$ to $C$ is a $\SPSP$ circuit of bounded formal degree $d $, bounded individual degree at most $r$, size $s$ and bottom support at most $t$. Using the same arguments as those in \cref{thm:fn-pow-d4}, we get that $C_V$ also functionally computes $P|_V(Y,Z)$. Thus,
\begin{align*}
  \sed(P|_V(Y,Z)) = \sed(C_V(Y,Z)).
\end{align*}
Further, from \cref{cor:psspdml-sed} and \cref{cor:sed-psspdr}, the above equation can be extended to the following inequality.
\begin{equation}
  \label{eqn:chain-ineq-D}
  \psspd(P|_V(Y,Z)) \leq \sed(P|_V(Y,Z)) = \sed(C_V(Y,Z)) \leq \psspdr(C_V(Y,Z)).
\end{equation}
From \cref{thm:psspd-chiexp}, we have that
\begin{equation}\label{eqn:fn-lb-D}
  \psspd(P|_V(Y,Z)) \geq \frac{\inparen{\frac{m}{m-\ell}}^{2\alpha k}\cdot {m-2\alpha k\choose \ell}}{2^{O(k)}\cdot\inparen{\frac{\ell}{m-\ell}}^{2\alpha k(1-\tau)}}
\end{equation}
and from \cref{lem:cktub}, we have that
\begin{equation}\label{eqn:fn-ub-D}
  \psspdr(C_V(Y,Z)) \leq s\cdot{\frac{2d }{t}+1\choose kr}\cdot{m\choose \ell+krt}\cdot{(\ell+krt)}.
\end{equation}

Putting \cref{eqn:chain-ineq-D}, \cref{eqn:fn-lb-D} and \cref{eqn:fn-ub-D} together, we get the following.
\begin{align*}
  \frac{\inparen{\frac{m}{m-\ell}}^{2\alpha k}\cdot {m-2\alpha k\choose \ell}}{2^{O(k)}\cdot\inparen{\frac{\ell}{m-\ell}}^{2\alpha k(1-\tau)}}\leq s\cdot {\frac{2d }{t}+1\choose kr}\cdot {m\choose \ell+krt}\cdot{(\ell+krt)}.
\end{align*}
Thus,
\begin{align*}
  s &\geq \frac{\inparen{\frac{m}{m-\ell}}^{2\alpha k}\cdot {m-2\alpha k\choose \ell}}{2^{O(k)}\cdot\inparen{\frac{\ell}{m-\ell}}^{2\alpha k(1-\tau)}}\times\frac{1}{{\frac{2d }{t}+1\choose kr}\cdot{m\choose \ell+krt}\cdot{(\ell+krt)}}\\
    &= \frac{\inparen{\frac{m}{m-\ell}}^{2\alpha k}}{2^{O(k)}\cdot {\frac{2d }{t}+1\choose kr}}\cdot\frac{(m-2\alpha k)!}{\ell!\cdot (m-\ell-2\alpha k)!}\cdot\frac{(\ell+krt)!(m-\ell-krt)!}{m!}\cdot\inparen{\frac{m-\ell}{\ell}}^{2\alpha k(1-\tau)}\\
    &= \frac{\inparen{\frac{m}{m-\ell}}^{2\alpha k}}{2^{O(k)}\cdot {\frac{2d }{t}+1\choose kr}}\cdot\frac{(m-2\alpha k)!}{m!}\cdot\frac{(m-\ell)!}{(m-\ell-2\alpha k)!}\cdot\frac{(\ell+krt)!}{\ell!}\cdot\frac{(m-\ell-krt)!}{(m-\ell)!}\cdot\inparen{\frac{m-\ell}{\ell}}^{2\alpha k(1-\tau)}\\
    &\approx \frac{\inparen{\frac{m}{m-\ell}}^{2\alpha k}}{2^{O(k)}\cdot {\frac{2d }{t}+1\choose kr}}\cdot\inparen{\frac{m-\ell}{m}}^{2\alpha k}\cdot\inparen{\frac{\ell}{m-\ell}}^{krt}\cdot\inparen{\frac{m-\ell}{\ell}}^{2\alpha k(1-\tau)}\\
    &= \inparen{\frac{m-\ell}{\ell}}^{2\alpha k(1-\tau) - krt}\cdot\frac{1}{2^{O(k)}\cdot {\frac{2d }{t}+1\choose kr}}\\
    &= \inparen{\frac{1+\Gamma}{1-\Gamma}}^{2\alpha k(1-\tau) - krt}\cdot\frac{1}{2^{O(k)}\cdot {\frac{2d }{t}+1\choose kr}}\\
    &\geq \inparen{(1+\Gamma)^2}^{2\alpha k(1-\tau) - krt}\cdot\frac{1}{2^{O(k)}}\\
    &= \inparen{(1+\Gamma)^\alpha}^{4k(1-\tau) - \frac{2krt}{\alpha}}\cdot\frac{1}{2^{O(k)}\cdot {\frac{2d }{t}+1\choose kr}}\\
    &\approx \inparen{2n^\varepsilon}^{4k(1-\tau) - \frac{2krt}{\alpha}}\cdot\frac{1}{2^{O(k)}\cdot {\frac{2d }{t}+1\choose kr}}\\
    &\geq {\inparen{{\Theta(1)}\cdot n^{4\varepsilon(1-\tau) - 10^{-2}}}^{k}}\inparen{\frac{krt}{e(2d +t)}}^{kr}\\
    &\geq n^{1.05k}\cdot\inparen{\frac{krt}{6d }}^{kr}
\end{align*}
In the above math block, in line 2 we absorb $(\ell+krt)$ into $2^{O(k)}$, in line 4 we use \cref{lem:bin-gkks} to get the approximations, in line 6 we use the fact that $m = \frac{\ell}{2}(1-\Gamma)$, in line -3 we use the fact that $(1+\Gamma)^\alpha \approx 2n^{\varepsilon}$, in line -2, we use the fact that $r$ is at most $\frac{\alpha}{200t}$, ${n\choose k}\leq \inparen{\frac{en}{k}}^k$ and in the last line we use the fact that $\varepsilon=0.29$ and $\tau=0.08$.

We shall now fix the values of $k$ and $t$ such that $t= k = \sqrt{\frac{d }{600r}}$. The above expression simplifies further to $s\geq \inparen{\frac{n^{1.05}}{({3600})^r}}^k$. If $r$ is at most $\frac{\log{n}}{12}\leq \frac{\log{n}}{\log{3600}}$, we get that $s\geq n^{0.05k}$. This setting of parameters also satisfies the criteria that $s\leq n^{\frac{t}{2}}$ and $r\leq \frac{\alpha}{200t}$. Under this substitution,
\begin{align*}
  r\leq \frac{\alpha}{200t}= \frac{1}{200k}\cdot \inparen{\frac{d}{2k}-\frac{1}{3}}\leq \frac{d}{401 k^2}\leq \frac{600r}{401}.
\end{align*}
\qed

\section{Acknowledgments}
The author is grateful to Nikhil Balaji, Mrinal Kumar, Noga Ron-Zewi, Nithin Saurabh, and Nithin Varma for helpful discussions. The author thanks Nikhil Balaji for telling him more about the Boolean complexity of Iterated Matrix Multiplication. The author thanks Ramprasad Saptharishi for patiently presenting the results in \cite{FKS16} while the author visited Tel Aviv University in 2016, hosted by Amir Shpilka.

\bibliographystyle{alphaurl}
\bibliography{ref}

\appendix

\section{Proof of \cref{lem:cktub}}
  Let $C$ be expressed as sum of terms $T_1+T_2+\ldots+T_s$ where each $T_i$ is a product of polynomials $Q_{i1}\cdot\ldots\cdot Q_{iD}$. W.L.O.G we can assume that all but one of the polynomials $Q_{i,j}$'s have a degree of at least $\frac{t}{2}\,.$ If not, pick two polynomials of degree strictly smaller than $\frac{t}{2}$ and merge them. Repeat this process until all but one of the factors have degree at least $\frac{t}{2}$. Note that for all $i\in [s]$, the formal degree of $T_i$ is at most the formal degree of $C$, and syntactic degree of the term $T_i$ is at most the formal degree of $T_i$. From the afore mentioned arguments, for all $i\in [s]$ syntactic degree of $T_i$ is at least $(D-1)\cdot\frac{t}{2}$ and formal degree of $T_i$ is at most $d $. Thus, $D$ is at most $\frac{2d }{t}+1$.

  From the sub-additivity of measure, we know that
  \begin{equation}
    \label{eqn:subadd-psspd}
    \psspd(C) \leq \sum_{i=1}^s\psspd(T_i).
  \end{equation}
  Let $T = Q_1\cdot\ldots\cdot Q_D$ be an arbitrary term in $\inbrace{T_1, \ldots, T_s}$. We shall henceforth obtain a bound on $\psspd(T)$ and then put it together with \cref{eqn:subadd-psspd} to get the desired result.

  
  We will first show by induction on $k$ the following for the set of $k$th order partial derivatives of $T$ with respect to degree $k$ monomials over variables from $Y$.
  \begin{align*}
    \partial^{=k}_Y T \subseteq &\fspan\inparen{\inbrace{\bigcup_{S\in{[D]\choose D-k}}\inbrace{\inparen{\prod_{i\in S} Q_i(Y,Z)}\cdot Z_{\inbrace{\leq kt}}\cdot \F[Y]}}}.
  \end{align*}
  The base case of induction for $k=0$ is trivial as $T$ is already in the required form. Let us assume the induction hypothesis for all derivatives of order $<k$. That is, $\partial^{=k-1}_Y T$ can be expressed as a linear combination of terms of the form 
  \begin{align*}
    h(Y, Z) = \inparen{\prod_{i\in S}Q_i(Y,Z)}\cdot h_1(Z)\cdot h_2(Y).
  \end{align*}
where $S$ is a set of size $D-(k-1)$, $h_1(Z)$ is a \emph{structured} polynomial in $\F[Z]$ such that $h_1(Z)$ can be expressed as a linear combination of multilinear monomials of support at most $(k-1)t$, and $h_2(Y)$ is some polynomial in $\F[Y]$. 

For some $u\in[\abs{Y}]$ and some fixed $i_0$ in $S$, 
\begin{align*}
  \frac{\partial h(Y,Z)}{\partial y_u} &= \inparen{\sum_{j\in S}\inparen{\prod_{\substack{i\in S\\i\neq j}}Q_i(Y,Z)}\cdot \frac{\partial Q_j(Y,Z)}{\partial y_u}\cdot h_1(Z)\cdot h_2(Y)}\\ &\qquad + \frac{\prod_{i\in S}Q_i}{Q_{i_0}}\cdot Q_{i_0}(Y,Z)\cdot h_1(Z)\cdot\frac{\partial h_2(Y)}{\partial y_u}
\end{align*}
where the first summand on the right hand side of the above equation lies in the subspace  $\fspan\inbrace{\inparen{\prod_{\substack{i\in S\\i\neq j}}Q_i(Y,Z)}\cdot \frac{\partial Q_j(Y,Z)}{\partial y_u}\cdot h_1(Z)\cdot \F[Y] ~:~ j\in[S]}$ and the second summand in the same equation, lies in the subspace $\fspan\inbrace{\frac{\prod_{i\in S}Q_i}{Q_{i_0}}\cdot Q_{i_0}(Y,Z)\cdot h_1(Z)\cdot \F[Y]}\,.$

Note that $\frac{\partial Q_j(Y,Z)}{\partial y_u}$ and $Q_{i_0}$ are polynomials such that every monomial in these depends on at most $t$ many variables from $Z$. Thus,
\begin{align*}
  \frac{\partial h(Y,Z)}{\partial y_u} &\in \fspan\inbrace{\bigcup_{T\in{S\choose \abs{S}-1}}\inbrace{\inparen{\prod_{i\in T}Q_i(Y,Z)}\cdot Z_{\inbrace{\leq t}}\cdot h_1(Z)\cdot \F[Y]}}.
\end{align*}
In the above expression, the contribution from the variables from $Y$, to the monomials in $\frac{\partial Q_j(Y,Z)}{\partial y_u}$ and $Q_{i_0}$ gets absorbed into $\F[Y]$ factor.

Recall the fact that $h_1(Z)$ is a linear combination of monomials of support at most $(k-1)t$. Thus, we get that, 
\begin{align*}
	\frac{\partial h(Y,Z)}{\partial y_u} &\in \fspan\inbrace{\bigcup_{T\in{[D]\choose D-k}}\inbrace{\inparen{\prod_{i\in T} Q_i(Y,Z)}\cdot Z_{\inbrace{\leq kt}}\cdot \F[Y]}}\,.
\end{align*}

From the discussion above we know that any polynomial in $\partial^{=k}_Y(T)$ can be expressed as a linear combination of polynomials of the form $\frac{\partial h}{\partial y_u}$. Further, every polynomial of the form $\frac{\partial h}{\partial y_u}$ belongs to the set 
\begin{align*}
	W &= \fspan\inbrace{\bigcup_{T\in{[D]\choose D-k}}\inbrace{\inparen{\prod_{i\in T} Q_i(Y,Z)}\cdot Z_{\inbrace{\leq kt}}\cdot \F[Y]}}\,.
\end{align*}
Thus, we get that $\partial^{=k}_YT$ is a subset of $W$. This completes the inductive argument.
	 
From the aforementioned discussion, we can now derive the following expressions.
\begin{align*}
  \sigma_Y\inparen{\partial^{=k}_Y T} \subseteq &\fspan\inbrace{\bigcup_{S\in {[D]\choose D-k}}\inbrace{\inparen{\prod_{\substack{i\in S}}\sigma_Y(Q_i)}\cdot Z_{\inbrace{\leq kt}}}}\,.
\end{align*}

It is easy to see that this inclusion holds under shift by monomials of degree at most $\ell$ over variables from $Z$.
\begin{align*}
  Z^{\leq \ell}\cdot\sigma_Y\inparen{\partial^{=k}_Y T} \subseteq &\fspan\inbrace{\bigcup_{S\in {[D]\choose D-k}}\inbrace{\inparen{\prod_{\substack{i\in S}}\sigma_Y(Q_i)}\cdot Z_{\inbrace{\leq \ell + kt}}}}\,.
\end{align*}
By taking a multilinear projection of the elements on both sides, we get that
\begin{align*}
  \fspan\inbrace{\mult\inparen{Z^{\leq \ell}\cdot \sigma_Y\inparen{\partial^{=k}_Y T}}}&\subseteq \fspan\inbrace{\bigcup_{S\in {[D]\choose D-k}}\inbrace{\mult\inparen{\inparen{\prod_{\substack{i\in S}}\sigma_Y(Q_i)}\cdot Z_{\inbrace{\leq \ell + kt}}}}}\\
  &\subseteq \fspan\inbrace{\bigcup_{S\in {[D]\choose D-k}}\inbrace{\inparen{\mult\inparen{\prod_{\substack{i\in S}}\sigma_Y(Q_i)}}\cdot Z^{\leq kt+\ell}_{\ml}}}.
\end{align*}
	
Thus we get that $\dim\inparen{\fspan\inbrace{\mult\inparen{Z^{\leq \ell}\cdot\sigma_Y(\partial^{=k}_Y T)}}}$ is at most 
\begin{align*}
  &\dim\inparen{\fspan\inbrace{\bigcup_{S\in {[D]\choose D-k}}\inbrace{\inparen{\mult\inparen{\prod_{\substack{i\in S}}\sigma_Y(Q_i)}}\cdot Z^{\leq kt+\ell}_{\ml}}}}\\
  \leq & \dim\inparen{\fspan\inbrace{\bigcup_{S\in{[D]\choose D-k}}\inbrace{\mult\inparen{\prod_{i\in S}\sigma_Y(Q_i)}}}}\cdot \dim\inparen{\fspan\inbrace{Z^{
	 \leq kt+\ell}_{\ml}}}\\
  \leq & {D\choose D-k}\cdot \sum_{i=0}^{kt+\ell}{m\choose i}\\
  \leq & {D\choose k}\cdot {m\choose \ell+kt}\cdot (\ell+kt) & \text{(Since $\ell+kt < m/2$)}.
\end{align*}

\end{document}